# Featured Articles /
## Articles de fond

## Citizen Science on the Faroe Islands in Advance of an Eclipse

*by Geoff Sims (geoffrey.sims@gmail.com)*
*& Kate Russo (umbraphillia@gmail.com)*

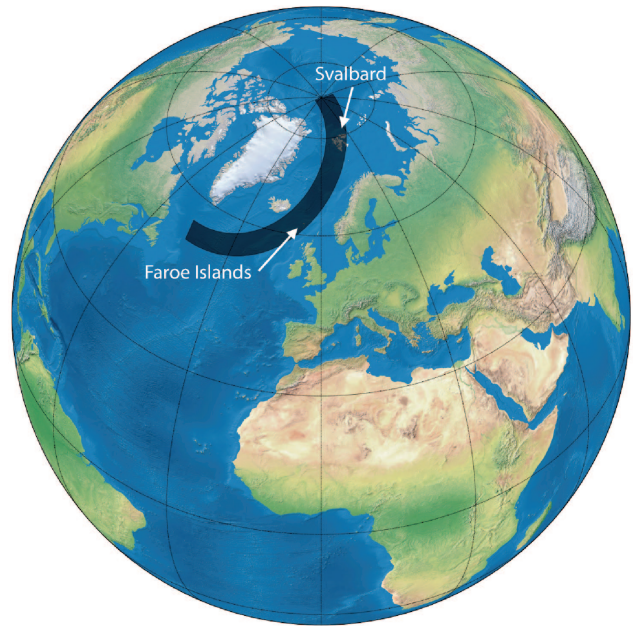

*Figure 1 — Path of the total solar eclipse on 2015 March 20.*

## Abstract

On 2015 March 20, a total solar eclipse will occur in the North Atlantic, with the Kingdom of Denmark's Faroe Islands and Norway's Svalbard archipelago (formerly Spitzbergen) being the only options for land-based observing. The region is known for wild, unpredictable, and often cloudy conditions, which potentially pose a serious threat for people hoping to view the spectacle.

We report on a citizen-science, weather-monitoring project, based in the Faroe Islands, which was conducted in March 2014—one year prior to the eclipse. The project aimed to promote awareness of the eclipse among the local communities, with the data collected providing a quantitative overview of typical weather conditions that may be expected in 2015. It also allows us to validate the usefulness of short-term weather forecasts, which may be used to increase the probability of observing the eclipse.

## Introduction

A total solar eclipse is one of the most amazing natural sights that one can witness. During this precise alignment of the Earth, Moon, and Sun, those fortunate enough to be situated in the path of the Moon's umbral shadow will observe daylight turn to eerie twilight darkness for a brief, few minutes. The total solar eclipse of 2015 March 20 will be visible from a narrow path traversing the North Atlantic and ending at the geographic North Pole (see the *Observer's Handbook 2015*, p. 126). The region of visibility includes only two landmasses: the Faroe Islands, and Svalbard. The total eclipse visibility path is shown in Figure 1.

As those who have seen a total solar eclipse will attest, the experience of totality can be described as euphoric, not to mention addictive. The eerie daytime darkness and elusive glimpse of the Sun's outer atmosphere—the corona—typically leaves one desperate to repeat the experience. More often than not this results in the self-diagnosed condition of "umbraphillia"—an addiction to being in the path of the Moon's shadow—and a personal goal to experience every subsequent total solar eclipse in one's lifetime.

The Faroe Islands' large permanent population (~50,000 people), excellent infrastructure, and relative ease of access will likely tempt thousands of people from around the world to choose these islands as their preferred viewing location in March 2015. We travelled to the Faroe Islands in March 2014, one year before the eclipse, to share information about the impending event. We spoke with school, tourism, and government authorities, and advised them on how best to plan for this celestial event.

During this visit, we undertook a "citizen science" project to monitor cloud conditions across the islands with an emphasis on the impact clouds would have on potential eclipse viewing. The project had several aims: to involve the local community in eclipse preparations; to collate snapshot observational weather data from across the islands; and to determine the usefulness of short-term weather forecasts.

This paper presents the results of this project; this section introduced the eclipse that will occur in March 2015, and gave a background on our visit to the islands last March. The next section gives an overview of the expected weather across the eclipse path. This is then followed by a description of the weather project, with the results quantitatively analyzed in the final section.

## Synoptic Situation

While it is currently possible to predict, with high precision, the paths of solar eclipses millennia in advance, there is one important factor that has the potential to prohibit the observation of eclipses, one that we cannot forecast with such accuracy: the weather. The North Atlantic is a cloudy and stormy place, no more so than in March, when it is subject to three major cloud-making influences: a humid, ice-free ocean surface; a cold winter/spring atmosphere that saturates easily; and a location near the main track of low-pressure systems that leave the North American continent and venture across the North Atlantic and Barents and Greenland Seas. Both the Faroe Islands and Svalbard are impacted by the passage of these migrating depressions, though Svalbard less so than the Faroe Islands.

While the Faroe Islands have their wintery days, their more southerly location embeds them in the main circulation of the Gulf Stream, and temperatures are much warmer than in Svalbard. The islands have an alpine-like climate characterized by wind, cloud, and moderate summer and winter temperatures. Because they are more exposed to the main track of North Atlantic lows, the weather is highly changeable from day to day, and strong winds with heavy rain can arrive at any time of the year.

Locally, weather on both the Faroe Islands and Svalbard is strongly influenced by the terrain, as there is a substantial up-and-down character to the landscape in both locations. In particular, the lee side of a mountain or high hill tends to have a little less cloud than upwind sites.

Climate statistics show that "clear" and "overcast" skies are rare, that there is a modest frequency of "few" and "scattered" clouds, and there is a high frequency of "broken" cloudiness (home.cc.umanitoba.ca/~jander; also see the *Observer's Handbook 2015*, p. 144). On average, this amounts to a mean cloudiness of 75 percent in the Faroe Islands.

These statistics are derived from observations made at Vagar airport, and it is unclear how these averages generalize to the rest of the archipelago. Anecdotally, the Faroes are known for having extremely variable weather, a trait that was observed first-hand during our March 2014 visit. It was not uncommon to have overcast skies and rain one minute, followed by clear, blue skies the next.

## Project Outline

The eclipse weather project was initiated with three main goals in mind. Primarily it was an outreach exercise, aimed to encourage local people to take an active interest in the sky conditions, creating an increased awareness of the eclipse and perhaps astronomy in general. Secondly, the data collected would provide an interesting overview of the weather—and in particular, the variability thereof—that may be expected to occur next year. Finally, it would provide a useful means to evaluate the reliability of short-term weather forecasts in the region.

In order to obtain a "snapshot" of the weather at a given time across the islands, volunteers around the islands were enlisted to make a photographic observation of the sky conditions each day at 9:40 a.m. (the time the total eclipse will occur) for the entire month of March 2014. If a photo was not possible, they simply recorded one of five categories that best described the cloud situation with respect to the visibility of the Sun:

1. mostly clear
2. partly cloudy (Sun visible)
3. thin cloud (Sun partially visible)
4. partly cloudy (Sun obscured)
5. overcast

The categories were chosen to best represent the impact on potential eclipse observing. Categories 1 and 2 would be typically unobstructed viewing; category 3 would be obstructed viewing, although not catastrophic; while in categories 4 and 5 the Sun would be completely obscured.

There were 15 volunteers in total. The majority consisted of staff from the municipal tourism offices, located in the following towns and cities: Fuglafjørður, Klaksvík, Runavík, Sandoy, Tórshavn, Tvøroyri, Vágar, and Vágur. Involvement of tourism representatives was highly advantageous from an outreach perspective, as they will likely be the points of contact for many visitors and tourists next year. A number of additional volunteers took part in the project from the following areas: Eiði, Tórshavn, Tvøroyri, Søldarfjørður, Strendur, Vestmanna, and Viðareiði. In lieu of an observer on the far western isle of Mykines, a Webcam image captured each morning was used to expand spatial coverage.

The geographical distribution of participants (including the Mykines Webcam), is shown on the map in Figure 2.

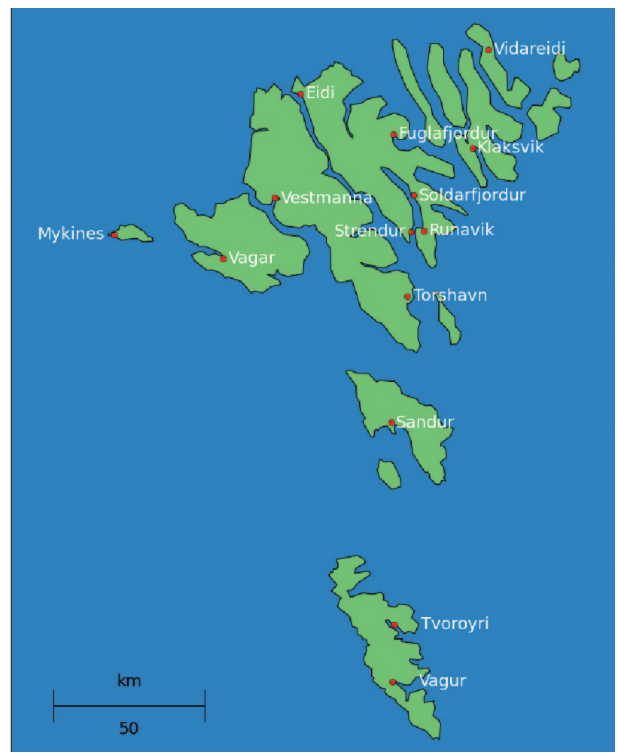

*Figure 2 — Geographic distribution of participants in the eclipse weather project.*

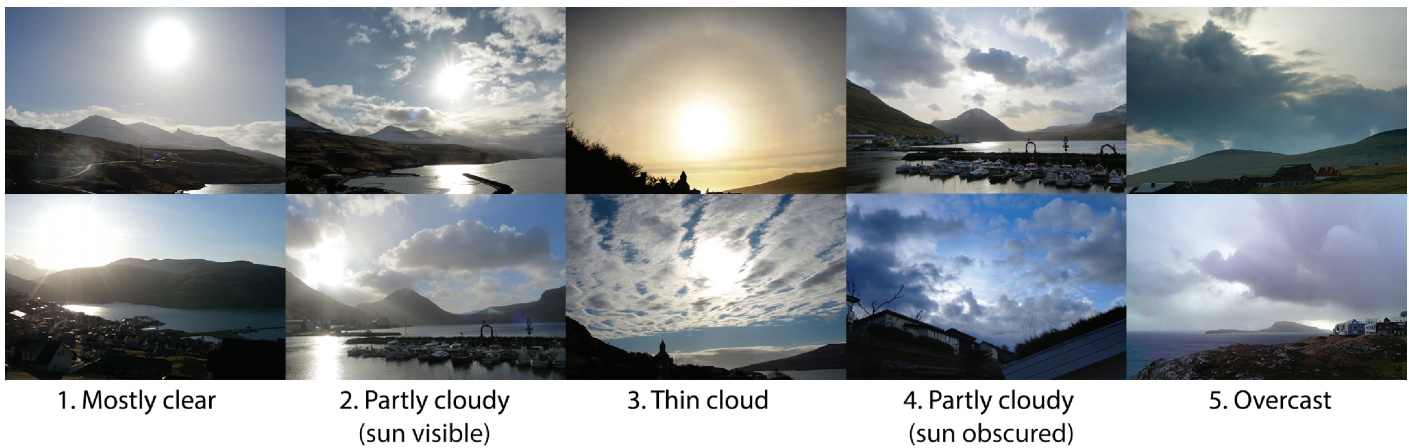

1. Mostly clear  2. Partly cloudy (sun visible)  3. Thin cloud  4. Partly cloudy (sun obscured)  5. Overcast

*Figure 3 — Sample selection of 10 photos across the 5 categories.*

In addition to these observations, every day for each location we obtained the 24-, 12-, 6-, and 3-hour forecasts from the Norwegian Meteorological Institute (www.yr.no, hereafter, YR), which included cloud estimates along with wind speed and direction. For additional verification purposes, Webcam images were captured at half-hour intervals every day from the Landsverk network (www.landsverk.fo) and elsewhere, which totalled 47 cameras around the Faroe Islands and 3 across Svalbard.

Each photograph was categorized (manually, by eye) into the five aforementioned categories, which allowed a quantitative analysis to be performed. To avoid any bias introduced by missing observations, nearby Webcam images captured at 09:30 were used as proxies (*e.g.* by looking at the sky or shadows on the ground to infer sunshine). If no suitably close Webcams were available, a best guess was made given the other observations obtained that day.

## Results

From a maximum of 465 observations that could have been submitted (15 observers for 31 days), we received a total of 381 reports (82 percent), of which 361 had corresponding photographs. Figure 3 shows a sample selection of photographs with their designated cloud categories.

Using the EXIF metadata stored in these photographs, we obtained the temporal distribution of observations, shown in the histogram in Figure 4. The sharp peak corresponds with the requested time for the images to be taken (09:40 UT). Approximately 85 percent of photographs were taken within 30 minutes of the nominal time.

As most photographs were taken with smart phones, it was assumed that the timestamps were automatically set by the mobile phone network and hence would be sufficiently accurate for our purposes. From digital cameras, it is likely the time stamps will have some error, and in some cases possibly be set to Daylight Saving Time (+1 hour).

It is noted that the spread of times at which observations were taken will introduce some inherent uncertainty, particularly given the extreme weather variability and its changes from minute to minute.

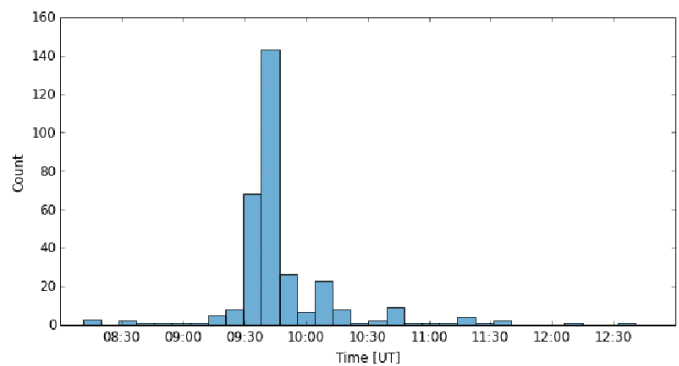

*Figure 4 — Histogram of observation times.*

The complete set of categorized observations, as well as the total cloud amount forecast six hours prior from YR, are presented in Figures 5 and 6.

- Only an extremely tiny fraction (2 percent) of reports were deemed "mostly clear," with "mostly cloudy" or "overcast" (categories 4 and 5) accounting for the vast majority of observations (Figure 6).

- Almost 50 percent of the time (15 days), at least one location was able to observe the Sun at the recorded time.

- There were 6 days (days 10, 13, 21, 22, 27, and 28), where conditions were "good" across most sites, although not one day existed where the entire archipelago experienced clear skies.

- The YR forecasts predicted most of the "good" observations, although on a number of occasions (days 2, 6, 20, and 29), clear skies were forecast in most areas, but not seen.

- A detailed inspection of Figure 5 provides a number of insights into the overall conditions, variability, and forecast

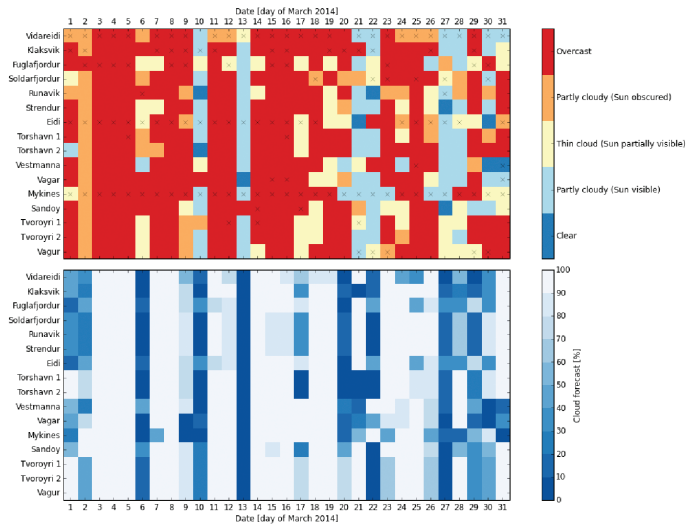

*Figure 5 — Grid representation of every categorized observation (upper panel) and forecast (lower panel) obtained. An "X" indicates that the category was estimated from a Webcam or nearby site. The colour scales on both plots are not the same.*

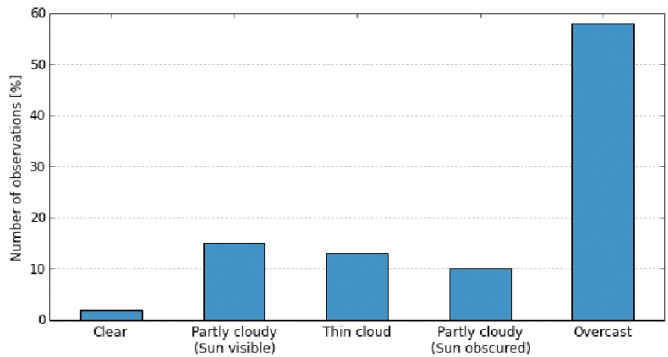

*Figure 6 — Percentages of observations in each of the five categories.*

accuracies within the month we examined, relating to the observed sky at 9:40 a.m.

- By looking at days that were cloudy everywhere except in one location (*e.g.* days 1, 6, and 24), there appears to be little correlation between site-specific forecasts and the actual observed conditions.

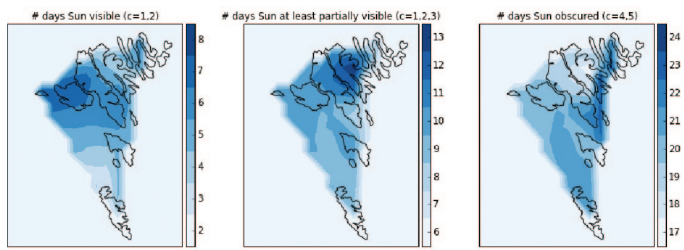

*Figure 7 — Spatial variation in observed weather. Note that in the left and middle plots, the darker blue represents a higher number of clear mornings, whereas in the right-most plot the dark colour represents a higher number of cloudy mornings.*

Despite having a sparsely spatially sampled dataset, it is possible to examine these aggregated statistics as a function of position. Figure 7 shows how the overall statistics varied over the islands, by using a natural-neighbour interpolation of the irregularly spaced data points to a regular grid, followed by a linear interpolation. The western isles of Vagar and Mykines appear to have the sunniest morning skies, while eastern locations generally experience a higher number of overcast mornings.

It is difficult to validate quantitatively the YR cloud forecasts, since they are averaged not only over an entire hour-long interval, but also over the entire sky. For each observation category, we have plotted the histograms of the forecast total cloud percentages (Figure 8), as obtained 6 hours prior to the observation. The histograms reiterate that while the forecasts are generally accurate, there were a number of occasions where cloudy forecasts had clear observations, and vice versa.

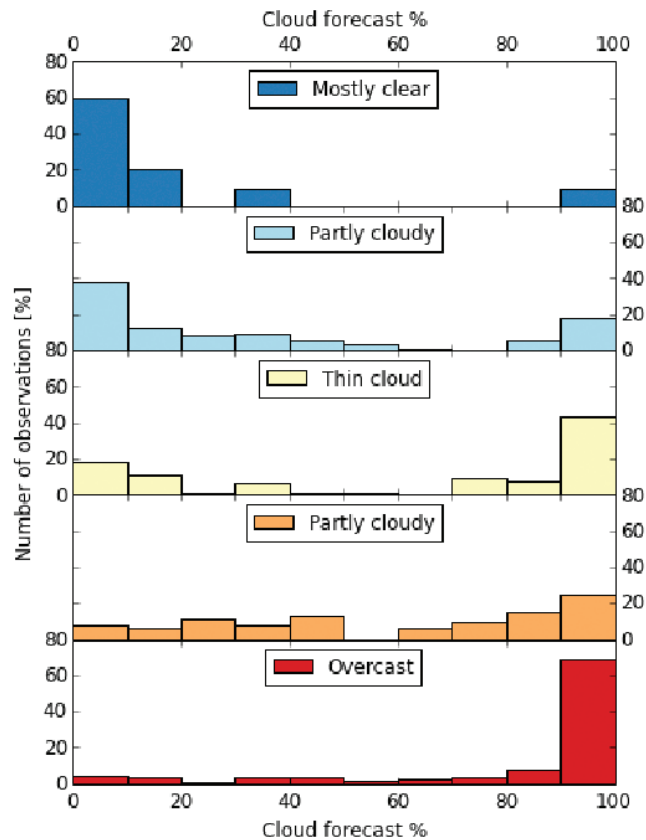

*Figure 8 — Histograms showing fractional amount of observations in each category as a function of forecast cloud percentage.*

Because the majority of photographs were classed in categories 4 and 5 (*e.g.* see Figure 6), the data in Figure 8 are hard to interpret from a forecast validation perspective. For example, although around 60 percent of all "clear sky" observations had cloud forecasts in the range 0–10 percent, if you consider every forecast within the range 0–10 percent, only 9 percent of those forecasts fell in the "clear sky" category (Figure 9).

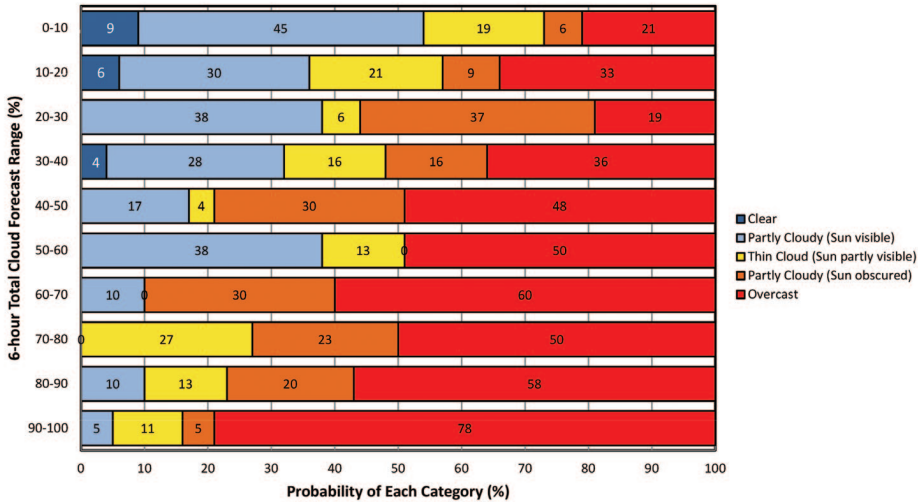

*Figure 9 — Probability of observing a particular sky condition, for a given 6-hour forecast.*

Figure 9 presents these statistics for all combinations, and is perhaps of most practical significance. That is, for a given forecast, it shows the probability of the actual sky conditions falling in any one of the five categories.

## Conclusions

The Faroese community is eagerly awaiting next year's solar eclipse. The involvement of the tourism industries and other civilians in this citizen science project has heightened awareness of the eclipse, weather, and the sky conditions in general.

From a meteorological perspective, the project has confirmed that, in general, the weather in the Faroe Islands is both highly variable and often cloudy. Even the best days usually have some cloud present, and on these exceptional days, it is likely some locations will still remain overcast. On a more positive note, based on this one month of data, we can conclude there is an almost even chance that somewhere on the islands clear skies will be experienced on the morning of the eclipse, and overall that there is around a 30-percent chance of not being completely clouded over.

Short-term (6-hour) cloud forecasts, in general, give a good prediction of the observed conditions. However, owing to the extremely variable nature of the weather, a significant proportion of the time is insufficiently accounted for by these forecasts. For example, 20 percent of forecasts indicating only 10–20 percent cloud were, in fact, observed to be overcast.

Despite the weather challenges that this eclipse presents, the Faroe Islands remain the destination of choice for many eclipse chasers, owing largely to the excellent tourism, communication, and road infrastructures, combined with easy accessibility via regular scheduled flights from Europe.

## Acknowledgements

We are extremely grateful to Visit Faroe Islands (www.visitfaroeislands.com), the Hafnia Hotel (www.hafnia.fo), and The Independent Traveller (www.independenttraveller.com) for supporting our visit. In particular, we thank Súsanna Sørensen, Tórstein Christiansen, and Rósa Remioff for their hospitality. We thank Sydney Observatory (www.sydneyobservatory.com) for their generous donation of eclipse glasses, which we distributed throughout the islands. Of course, this project would not have been possible without the cooperation of the municipality offices and associated volunteers: Jóannis Sørensen, Birna Joensen, Petur Petersen, Arnfríð Højgaard, Alma Poulsen, Elin Henriksen, Beinta Olsen, Lis Mortensen, Tórstein Christiansen, Andreas Joensen, Jóngerð Mikkelsen, Hanna Reynatúgvu, Póla Strøm, Turið Vestergaard, and Majken Gaard. This work makes use of the Natural Earth base map (naturalearthdata.com) and the *Python* programming language (www.python.org), including the *numpy* (www.numpy.org) and *matplotlib* (matplotlib.org) modules. ✶